\documentclass{acm_proc_article-sp}

\usepackage{multirow}
\usepackage{array}
\usepackage{subfigure}
\usepackage{scrextend}
\usepackage{threeparttable}
\usepackage{hyperref}

\title{When is it Biased? Assessing the Representativeness of Twitter's Streaming API}
\numberofauthors{3} 
\author{
\alignauthor
Fred Morstatter\\
       \affaddr{Arizona State University}\\
       \affaddr{699 S. Mill Ave}\\
       \affaddr{Tempe, AZ, 85281}\\
       \email{fred.morstatter@asu.edu}
\alignauthor
J{\"u}rgen Pfeffer\\
       \affaddr{Carnegie Mellon University}\\
       \affaddr{5000 Forbes Ave.}\\
       \affaddr{Pittsburgh, PA, 15213}\\
       \email{jpfeffer@cs.cmu.edu}
\alignauthor 
Huan Liu\\
       \affaddr{Arizona State University}\\
       \affaddr{699 S. Mill Ave}\\
       \affaddr{Tempe, AZ, 85281}\\
       \email{huan.liu@asu.edu}
}

\begin{document}
\maketitle

\begin{abstract}
  \begin{quote}
  Twitter has captured the interest of the scientific community not only for its massive user base and content, but also for its openness in sharing its data. Twitter shares a free 1\% sample of its tweets through the ``Streaming API'', a service that returns a sample of tweets according to a set of parameters set by the researcher. Recently, research has pointed to evidence of bias in the data returned through the Streaming API, raising concern in the integrity of this data service for use in research scenarios. While these results are important, the methodologies proposed in previous work rely on the restrictive and expensive Firehose to find the bias in the Streaming API data. In this work we tackle the problem of finding sample bias without the need for ``gold standard'' Firehose data. Namely, we focus on finding time periods in the Streaming API data where the trend of a hashtag is significantly different from its trend in the true activity on Twitter. We propose a solution that focuses on using an open data source to find bias in the Streaming API. Finally, we assess the utility of the data source in sparse data situations and for users issuing the same query from different regions.
  \end{quote}
\end{abstract}



\section{Introduction}
Twitter is a microblogging site where users share short, 140-character messages called ``tweets''. Twitter has become one of the largest social networking sites in the world with 240 million users who publish 500 million tweets each day\footnote{\url{https://blog.twitter.com/2013/new-tweets-per-second-record-and-how}} from their web browsers and mobile phones. Due to Twitter's massive size and ease of mobile publication, Twitter has also become a central tool for communication during protests~\cite{Camp2011} and disasters~\cite{human-un}. This has caused an immense push from both the computer and social science research communities who have used data from the site for wide applications of social media research from predicting users' location~\cite{Chen10} or to characterize the life cycle of news stories~\cite{castillo2013characterizing}.

Twitter's policy for data sharing is very open, providing a free ``Streaming API''\footnote{\url{https://dev.twitter.com/docs/streaming-apis}} that returns tweets matching a query provided by the Streaming API user. One drawback of the Streaming API is that it only returns at most 1\% of the tweets on Twitter at a given moment. Once the volume of the query surpasses 1\% of all of the tweets on Twitter, the response is sampled. The way in which Twitter samples the data is unpublished. Recent research~\cite{Morst-etal13} has shown that there is evidence of bias in this sampling mechanism under certain conditions, leaving researchers to wonder \emph{when} the Streaming API is representative, and when it is biased. 

One way to get around the 1\% limit is to purchase the Twitter Firehose, a feed offered by Twitter that allows for access to 100\% of all of the public tweets posted on the site. Simply comparing the results from the Streaming API with the Firehose is one way to verify the results of the sampled service. This is the approach taken in~\cite{Morst-etal13}. Unfortunately, verifying  results using the Firehose is not an option for most researchers as access to Twitter's Firehose is restrictively expensive and access is limited to users with a business agreement with Twitter\footnote{\url{https://dev.twitter.com/discussions/2752}}. 

In this work, we define ``bias'' as sample bias. We say that a hashtag is ``biased'' if the relative trend is statistically significantly over-represented or underrepresented in contrast to its true trend on Twitter. In particular, we are looking for particular time periods of bias in the Streaming API. Based on Twitter's documentation, the sample size is determined by the volume of a query at a point in time. There are times when the sample is representative, and times when it is biased. We try to find time periods where the data from the Streaming API is biased, meaning not an accurate representation of the true activity on Twitter. 

The focus of this paper lies in finding and assessing an alternative method for bias detection in the Streaming API. Our goal is to detect the bias automatically, using methods that are openly available to researchers. We begin this process by discussing the related work, which includes a discussion of other work that discovers bias in the Streaming API. We continue to introduce and vet another data source, the Sample API\footnote{\url{https://dev.twitter.com/docs/api/1/get/statuses/sample}}, and propose a methodology that utilizes this data source to show a user when there is bias in the results of their Streaming API query. Finally, we show that the Streaming API gives nearly the same results to identical queries originating from different points around the world, and to identical queries started at different points in time during the overlap of their execution. This work is the full version of~\cite{Morst-etal14}.

\section{Related Work}
Our related work falls into three areas: one concerning work previously done on Twitter's Streaming API, another concerning research done to verify other black box systems on the web, and finally another area devoted to previous evidence of bias found on Twitter's Streaming API.

\subsection{Work with the Streaming API}
Twitter's Streaming API has been used throughout the domain of social media and network analysis to generate understanding of how users behave on these platforms. It has been used to collect data for topic modeling~\cite{Hong-Davi10,Pozd-11}, network analysis~\cite{Sofe12}, and statistical analysis of content~\cite{Math10}, among others. Researchers' reliance upon this data source is significant, and these examples only provide a cursory glance at the tip of the iceberg. 

\subsection{Bias in Black Box Systems}
The topic of assessing the results from a black box system is related to our work. Another web sciences black box that has been studied is Amazon's Mechanical Turk (AMT). In~\cite{crump2013evaluating}, the authors assess the representativeness of the users on AMT, and provide example tasks showing areas where these users give good performance. In a similar vein,~\cite{snow2008cheap} proposes a method to correct for response bias in the results obtained from AMT. In the area of social media research, this topic has also been studied from the perspective of link propagation. In~\cite{de2010does}, the authors analyze the effect data sampling has on the way link propagation is perceived. Specifically, the authors study URLs shared on sampled Twitter data.

\subsection{Bias in Twitter Data}
There are many potential areas of bias in Twitter. One possible area of bias in Twitter data comes from the demographic makeup of the users on the site. In~\cite{mislove2011understanding}, the authors use last names of users to estimate their race, and find that the ethnicity/race distribution of Twitter users diverges widely from U.S. Census estimates.~\cite{duggan2013demographics} finds similar results, and also finds that adults aged 18-29, African Americans, and urban residents are over-represented on the site.

The bias we focus on in this work is concerned with sample bias from Twitter's APIs. The work performed in~\cite{Morst-etal13}~compared four commonly-studied facets of the Streaming API and Firehose data, looking for evidence of bias in each facet. They obtained widely different results across facets. First, they studied the statistical differences between the two datasets, using correlation to understand the differences between the top $n$ hashtags in the two datasets. They find some bias in the occurrence of the top hashtags for low values of $n$.

The authors also compared topical facets of the text by extracting topics with LDA~\cite{blei-etal03}, where they found similar evidence of bias. The authors discover that the topics extracted through LDA are significantly different than those extracted from the gold standard Firehose data. The other facets compared in the data were the networks extracted from the dataset. Here, the authors extracted the User $\times$ User retweet network from both sources and compare centrality measures across the two networks. They find that, on average, the Streaming API is able to find the most central users in the Firehose 50\% of the time. The final facet they compare is the distribution of ``geotagged'' tweets. Here, they find no bias and that the number of geotagged tweets from the Streaming API is over 90\% of that in the Firehose.

\section{Discovering Bias in the Streaming API without the Firehose}
With work showing evidence that the Streaming API is biased, researchers must be able to tell whether their Streaming API sample is biased. Vetting their dataset using the methodology proposed with the Firehose is prohibitive for many reasons. We propose a methodology that can give an indication of bias for a particular hashtag.

In this section, we investigate whether another open data source, Twitter's Sample API, can be used to find bias in the Streaming API. We show that using the Sample API, one can accurately detect bias in the Streaming API without the need of the prohibitive Firehose. We focus on alternative methods to help the user understand \emph{when} their data diverges from the true activity on Twitter. We continue to show that not only can one find the bias using this method, but that these results are consistent regardless of when and where the Streaming API query was issued.

\subsection{Validation of the Streaming API}
We define the true trend of a particular hashtag, $h$ as a function, $f(h)$. We define the trend of $h$ as it is conveyed through the Streaming API as $t(h)$. To make this estimation, we consider another freely-available open Twitter data source, the Sample API. Unlike the Streaming API, the Sample API takes no parameters and returns a 1\% sample of all of the Tweets produced on Twitter. 
Here, we empirically assess the Sample API's ability to return a truly random sample and continue to build a framework to compare the Streaming API data with the Firehose using the Sample API as a proxy. We call the trend from the Sample API $s(h)$. Given the sparsity of the Sample API, it is likely impossible to find the real trend $f(h)$ from $s(h)$, however $s(h)$ can be used as an indicator to understand when $t(h)$ is biased.

\begin{table}
  \centering
  \caption{Data Collected to Test Bias Detection}
  \begin{tabular}{|c|c|c|c|}
  \hline
  Data Source & Keywords & No. Tweets \\
  \hline
  Firehose (Gnip) & syria & 214,383 \\
  \hline
  Sample API & N/A & 734,172 \\
  \hline
  \hline
  \end{tabular}
  \label{tab:moredata}
\end{table}

\begin{table}
  \centering
  \caption{Significance levels of $\tau_\beta$ statistic for top $k$ hashtags, Sample API VS. Firehose.}
  \begin{threeparttable}
    \begin{tabular}{| c | c | c |}
      \hline
      Top-$k$ & $\tau_\beta$ & $p$-value \\
      \hline
      10 & 0.988826 & 0.000069 \\
      \hline
      20 & 0.778663 & 0.000001 \\
      \hline
      30 & 0.655072 & 0.000001 \\
      \hline
      40 & 0.549759 & 0.000002 \\
      \hline
      50 & 0.604880 & $< 10^{-6}$ \\
      \hline
      ... & ... & ... \\
      \hline
      450 & 0.476931 & $< 10^{-6}$\tnote{*} \\
      \hline
      \hline
    \end{tabular}
    \begin{tablenotes}
    \item [*] All lists of size greater than 40 had $p$-values $< 10^{-6}$.
  \end{tablenotes}
  \end{threeparttable}
  \label{tab:kt}
\end{table}

\subsection{Vetting the Randomness of the Sample API}
Given the evidence of bias that was observed in the Streaming API, we must proceed with caution before using the Sample API as a surrogate gold standard. We begin our assessment of the randomness of the Sample API by collecting data from the feed. We collect all of the tweets available through the Sample API on 2013-08-30 from 17:00 - 21:00 UTC, and post-filter them by the keyword ``syria''. Simultaneously, we collect all of the tweets matching the keyword ``syria'' from the Gnip Twitter feed. The Gnip\footnote{\url{http://gnip.com/}} feed is another outlet for Firehose data, it also provides 100\% of the publicly-available Tweets on Twitter. We report the keywords and the collection volume for this dataset in Table~\ref{tab:moredata}.  

\begin{figure}
\includegraphics[width=0.5\textwidth]{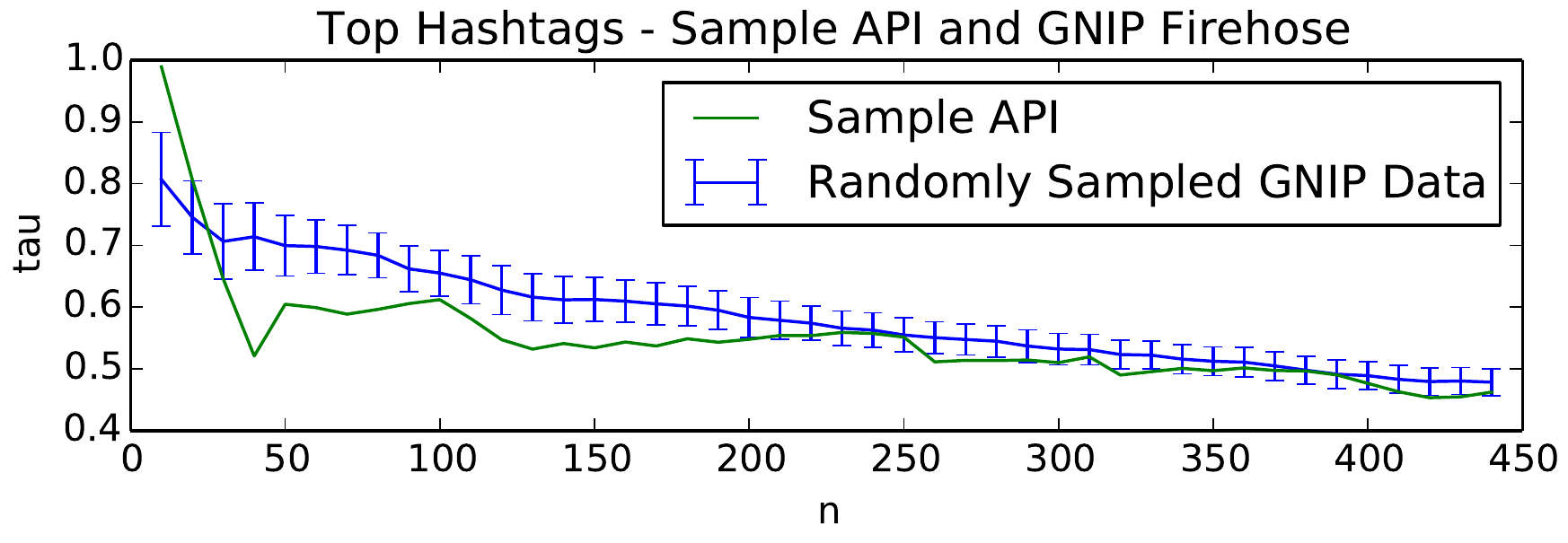}
\caption{Rank correlation of Sample API and Gnip Firehose. Relationship between $n$ - number of top hashtags, and $\tau_\beta$ - the correlation coefficient for different levels of coverage.}
\label{fig:samplen}
\end{figure}

To verify the validity of this source we compare the ranked list of top hashtags in both sets. We first plot Kendall's $\tau_\beta$ score of the Sample API against the Firehose. Kendall's $\tau_\beta$ calculates the number of concordant and discordant pairs between two ranked lists. This score gives us a sense of whether the frequency of hashtags coming through the Sample API is the same as the Firehose. We plot the average and standard deviation of 100 perfectly randomly sampled datasets of the same size as the Sample API against the Firehose. A plot of the rank correlation in the top hashtags from the Firehose and the Sample API is shown in Figure~\ref{fig:samplen}. 

Overall we see that the shape of the trend of the Sample API closely resembles that of the random samples, a promising sign that the Sample API is unbiased. However, we still see that the $\tau_\beta$ values occasionally fall outside of one standard deviation of the distribution of random samples. We perform a statistical test to ensure that the Sample data and Firehose data are not independent. To perform the statistical test we calculate two-sided significance level from the Kendall $\tau$ statistic, which tests the following hypothesis:
\begin{addmargin}[2em]{2em}
$H_0$ -- The top $k$ hashtags in the Firehose data and the top $k$ hashtags in the Sample API data are independent, $\tau_\beta$ = 0.
\end{addmargin}

The results of this experiment at varying levels of $k$ are shown in Table~\ref{tab:kt}. In all cases we are able to reject $H_0$ with a 95\% confidence level. Given the strong correlation between the top hashtags and the strong similarity between the two distributions, we go forward with the understanding that the top hashtags from the Sample API are representative of the true activity on the Firehose.

With the revelation that the Sample API is a random sample of the Firehose, one might be tempted to simply use this as a replacement for the Streaming API, post-filtering the Sample API's data to suit their needs. While this is correct from the perspective of data sampling, this approach will lead to a massive loss in data. For example, assume that a query matches 10\% of the data on the Firehose. Filtering the Sample API will give us a dataset that is 0.1\% of the size of the Firehose, but unbiased. The Streaming API's sometimes-biased sampling method will give us a 1\% sample of the Firehose that is 10x larger than the Sample API. This may be counterintuitive as larger samples have a greater chance of being unbiased, however since the issue resides in the sampling methodology of the Streaming API a sample of any size from this source has the potential to be biased. The size difference between the two sources cannot be ignored. Instead, we will use the Sample API to identify periods of bias in the Streaming API.

\begin{figure*}[t]
     \begin{center}
        \subfigure[Streaming API Results. Trendline for ``\#believemovie'' over one day.]{%
           \label{fig:steps_streaming}
           \includegraphics[width=0.45\textwidth]{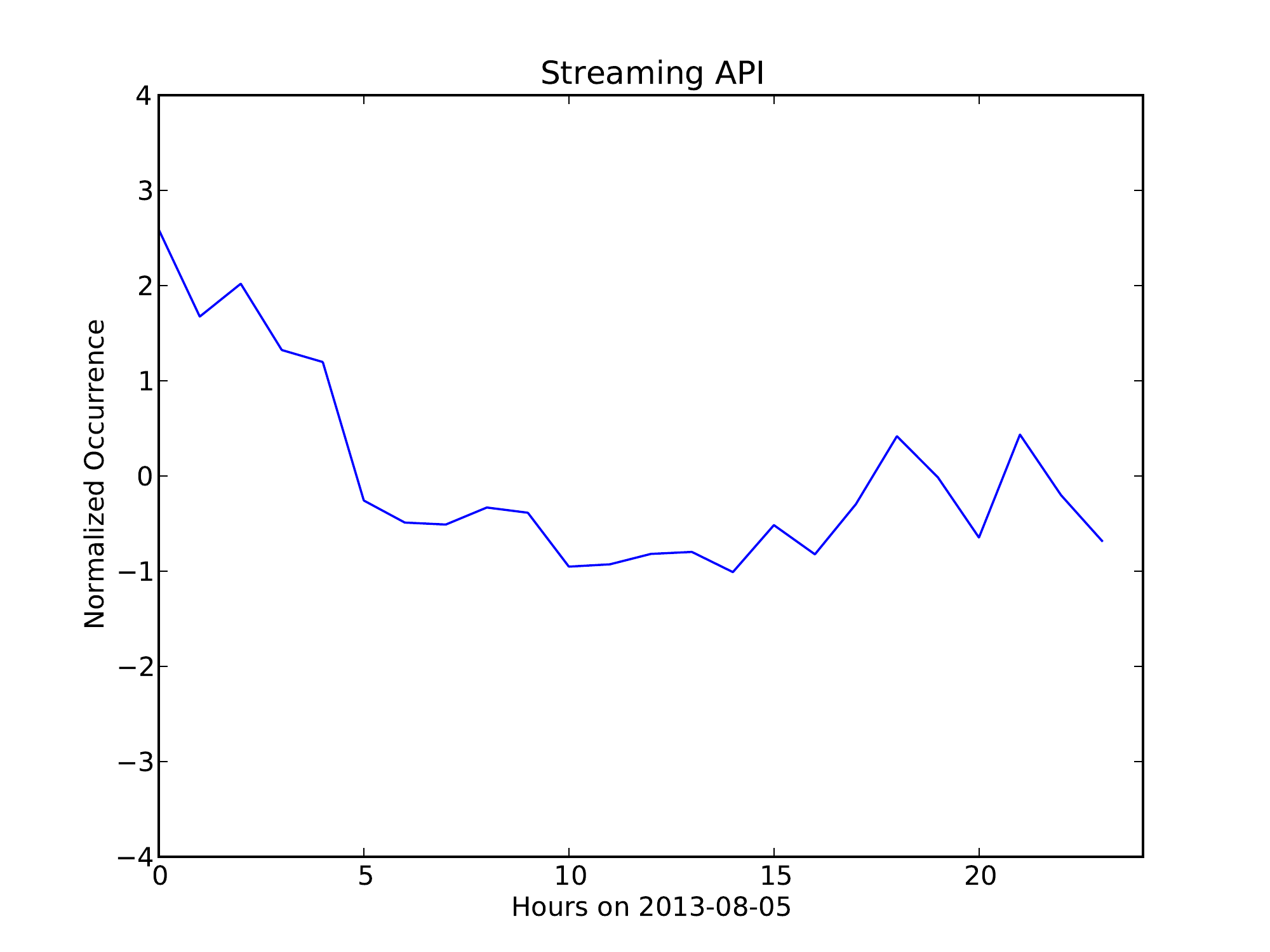}
        }~ %
        \subfigure[Sample API Results. Trendline for ``\#believemovie'' over one day.]{
            \label{fig:steps_sample}
            \includegraphics[width=0.45\textwidth]{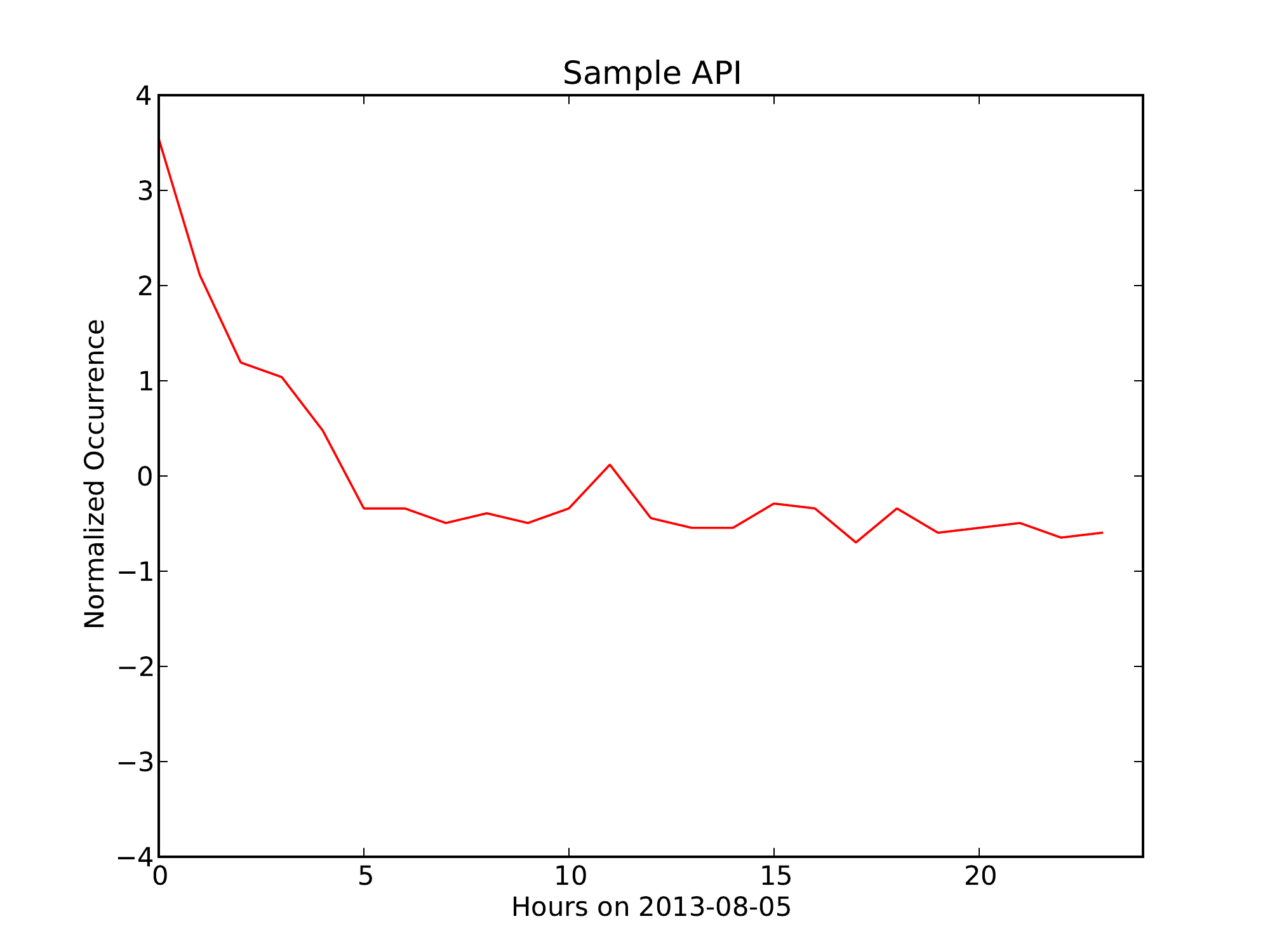}
        }\\%
        \subfigure[Trendlines from 100 bootstrapped samples of the  Sample API.]{%
           \label{fig:steps_bs_samples}
           \includegraphics[width=0.45\textwidth]{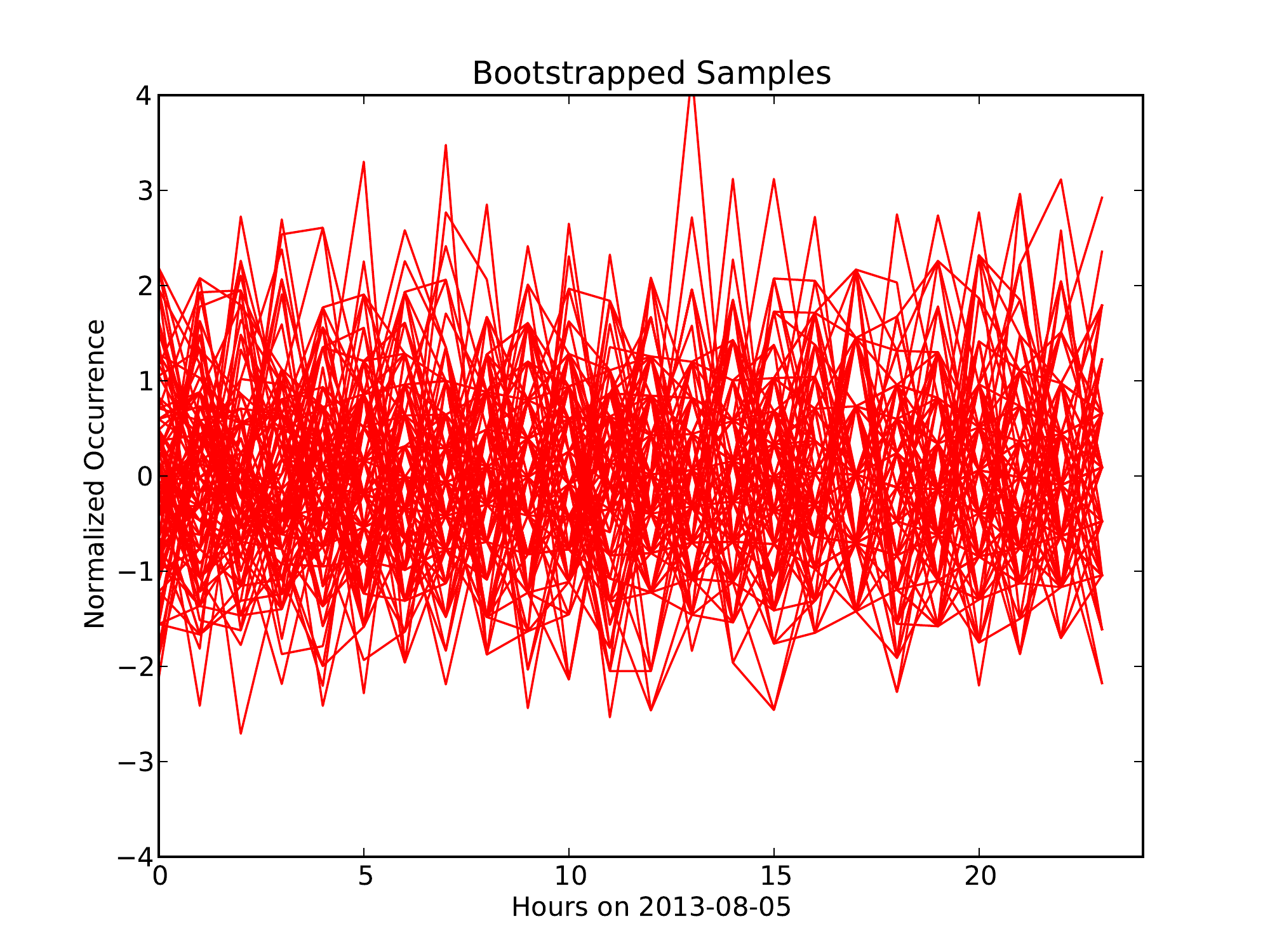}
        }~ %
        \subfigure[Bootstrapped average and $\pm3$ standard deviations overlaid with Streaming API.]{
            \label{fig:steps_sample_streaming_overlay}
            \includegraphics[width=0.45\textwidth]{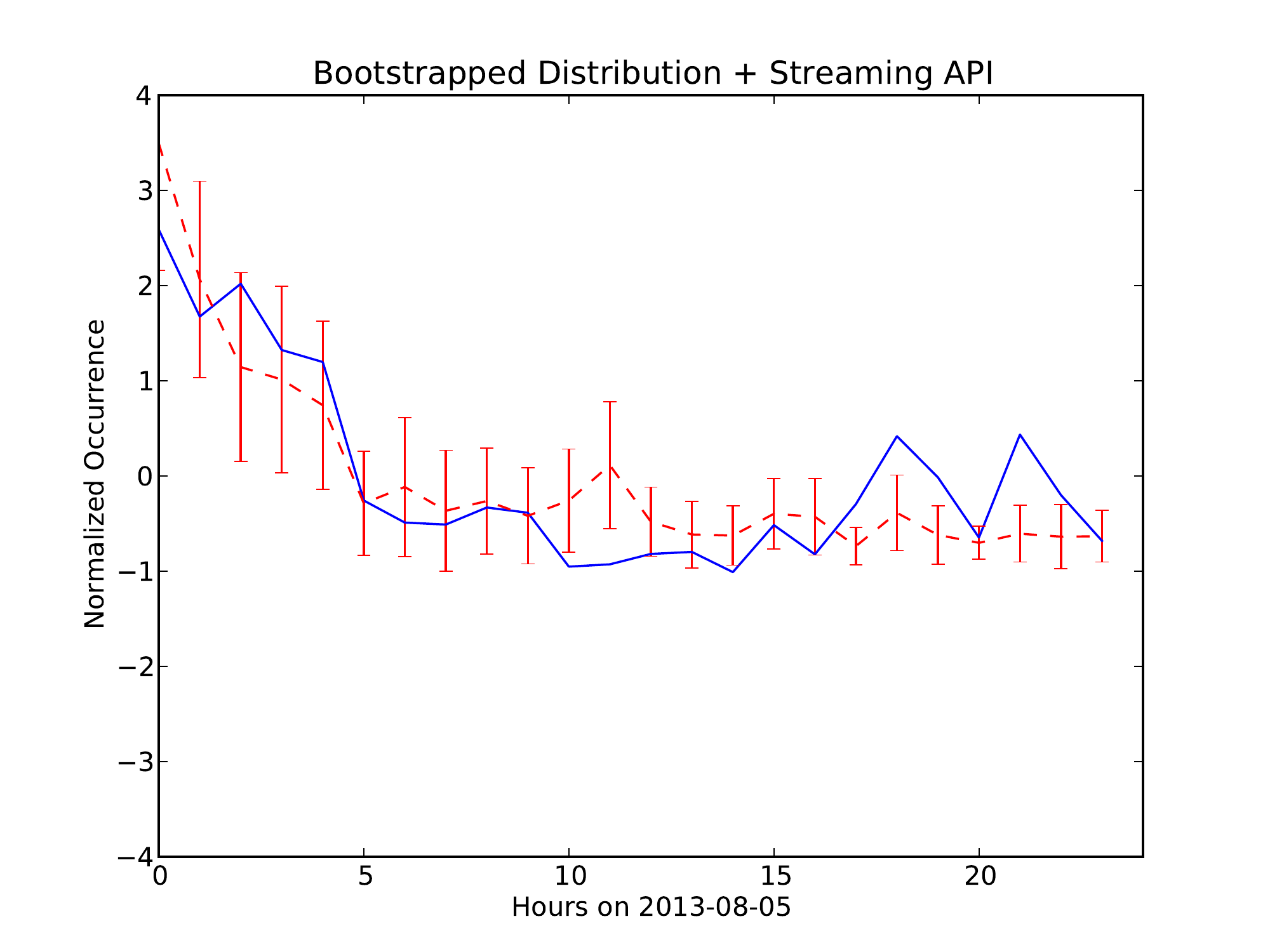}
        }%
    \end{center}
    \caption{Figures outlining different steps of the process for finding bias in the Streaming API.}
   \label{fig:overall_process}
\end{figure*}

\subsection{Finding Bias in the Trend of a Hashtag} 
Now that we know that the Sample API gives us an unbiased picture of Twitter's Firehose, we can continue to construct a framework that incorporates this source to find bias. Herein, we propose a methodology that finds the bias in the Streaming API and reports to the user collecting the data when there is likely bias in the data.

With only one unbiased view from the Sample API, it is difficult to understand what we should expect from our Streaming API data. When the results from both sources match there is clearly no problem. When there is a difference how do we know if the relative error between the Sample API and the Streaming API at one time step is significant or if it is just a small deviation from a random sample? To better understand the Sample API's response, we bootstrap~\cite{efron1982jackknife} the Sample API to obtain a confidence interval for the relative activity for the hashtag at a given time step.

We begin by normalizing both the Sample API and Streaming API time series. This is done by calculating the mean, and standard deviation of each of the counts in the time series. Finally, we normalize each point by its standard score, which is calculated as:
\begin{equation}
  Standard\_Score(t_i) = \frac{t_i - \mu_T}{\sigma_T},
\end{equation}
where $\mu_T$ and $\sigma_T$ are the mean and standard deviation of all of the time periods in the time series, respectively, and $t_i$ is an individual time series point. This is done to ensure that the distribution of points from both time series is $\mathcal{N}(0,1)$. We create 100 bootstrapped samples for each hashtag. We then extract the time series data from each sample and normalize them as we did before. This gives us a distribution of readings for each time period in the dataset. Next, we compare this distribution to the normalized time series from the Streaming API to detect the bias. We take the sample mean and sample standard deviation of this distribution at each point $t_i$ as $\mu^b_i$ and $\sigma^b_i$. Borrowing the threshold used in control charts~\cite{ryan2011statistical}, we say that any Streaming API value at time $t_i$ that is outside of $\pm3 \sigma_{t_i}$ is biased.

We show a full example of our method in Figure~\ref{fig:overall_process}. We enumerate the process for a single hashtag, ``\#believemovie'' on August 5th, 2013. We choose this hashtag as it is one of the most frequent hashtags on this day. The process begins with the time series data for this hashtag from both the Streaming and Sample APIs, shown in Figures~\ref{fig:steps_streaming} and~\ref{fig:steps_sample}, respectively. Looking at the two figures, we immediately see a difference in the trends of this hashtag from the two sources. To obtain a confidence interval on the difference of the two sources, we create 100 bootstrapped samples. The time series extracted from these samples are shown in Figure~\ref{fig:steps_bs_samples}. Finally, taking the mean and 3 standard deviations of the bootstrapped samples at each time point, we obtain the confidence intervals seen in Figure~\ref{fig:steps_sample_streaming_overlay}. We make several observations from Figure~\ref{fig:steps_sample_streaming_overlay}. First, a spike that occurs between hours 10 and 11 is underrepresented in the Streaming data. Also, the spikes that appear after hour 16 are both over-represented in the Streaming API. Due to the nature of our bootstrapping method, these observations are all statistically significant at the 99.7\% confidence interval.

\subsection{Signal Usability Under Sparsity}
One potential drawback of our method lies in the sparsity of the Sample API. Accounting for only 1\% of the Firehose, the ``long tail'' of hashtags will largely be ignored by the Sample API. This is problematic for researchers who wish to verify their Streaming API query's results when their query is focused upon hashtags that do not see a lot of activity as a fraction of the entire activity on Twitter. One counterargument is that these kinds of queries will likely not eclipse the 1\% threshold, and in general will be unbiased. 

One observation we make is that the times where our bootstrapping method will be futile is in times where there is data for the hashtag from the Streaming API and no data from the Sample API. In such cases, a bootstrapping approach will give us a degenerate distribution with mean 0, not allowing for meaningful comparison between the sources. We test the sparsity of the Sample API by finding ``known zeros'', hashtags seen in the results of the Streaming API query, but not in the Sample API for a particular time unit.

\begin{figure}
  \includegraphics[width=0.5\textwidth]{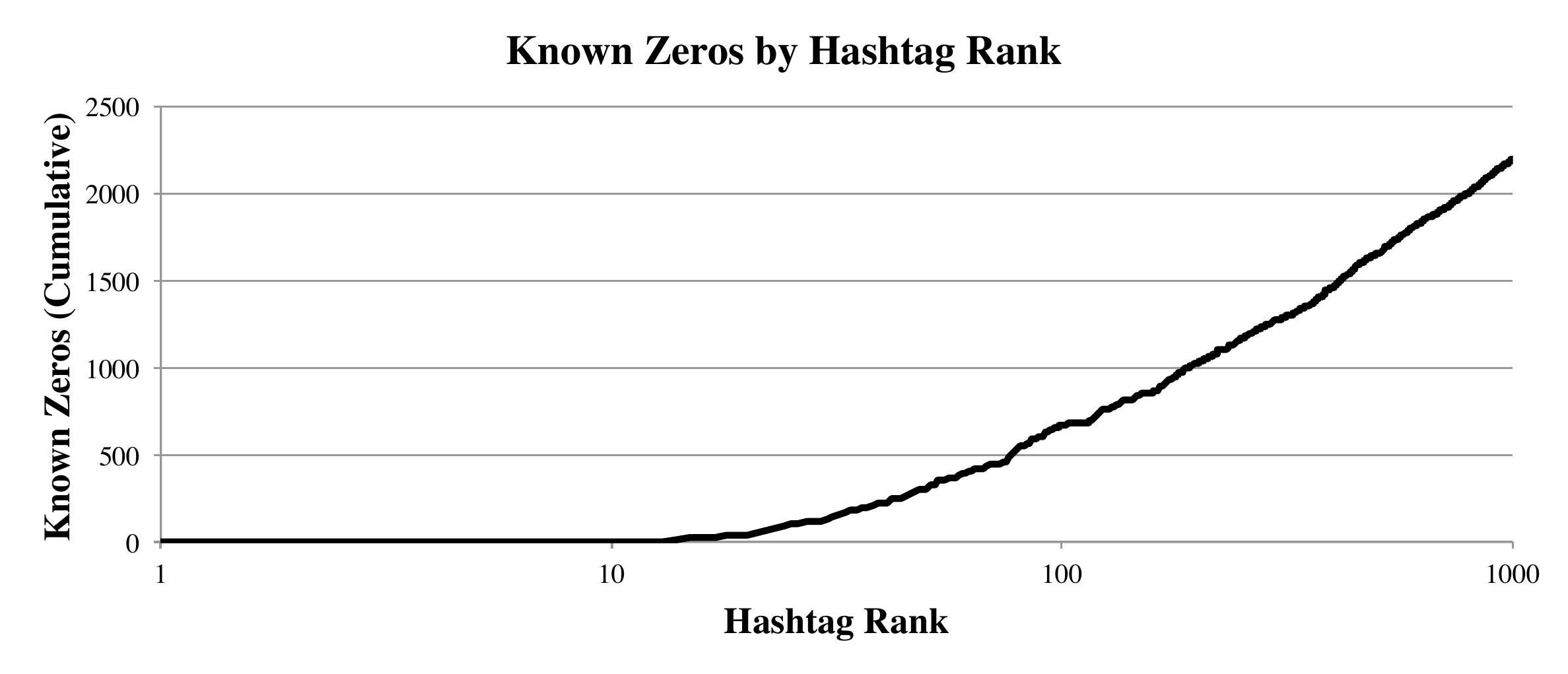}
  \caption{Cumulative known zeros ranked by hashtag popularity. We see that the most popular hashtags have relatively few points of missing data, while the less popular hashtags have many more. To help the reader separate the higher-ranked hashtags' missing values, we plot the $x$-axis on a log scale.}
  \label{fig:zeros}
\end{figure}

Figure~\ref{fig:zeros} shows the number of known zeros in the top 1,000 hashtags, with each hashtag ordered by frequency. Here, we see that the first hashtags are nearly perfect, with a total of only 4 known zeros in the top 10 hashtags. However, as we continue down the list, we begin to see more and more known-zeros. While this method helps researchers to find bias in their Streaming API queries, there are still many hours for many hashtags where no claim can be made about the validity of the data.



\section{Geographic and Temporal Stability of Queries}
In addition to tackling the bias problem, we also analyze the stability of the results when the data are collected in different geographic areas, and for queries started at different times. Do identical Streaming API queries started at different times get different responses during the overlap of their execution? Do identical Streaming API queries get different responses if they are executed from different geographical regions? To ensure that the results obtained in this paper hold for researchers outside of the US, we assess whether Twitter issues the same results to identical queries. 

To answer these questions, we collected data from the Streaming API in both the United States (USA) and Austria (AT) with the following scheme: every 20 minutes, we start a query that lasts for 30 minutes. For example, $query_1^{USA}$ and $query_1^{AT}$ collect tweets from 00:00 - 00:30 UTC, $query_2^{USA}$ and $query_2^{AT}$ from 00:20 - 00:50 UTC, and $query_3^{USA}$ and $query_3^{AT}$ from 00:40 - 01:10 UTC, and so on. Each query is configured with exactly the same parameters. In structuring our queries this way, we can control both for time, and for location. By looking at the 10-minute overlaps in the adjacent within-country queries (i.e. all $query_i^{C}$ and $query_{i+1}^{C}$), we can gain an understanding of whether identical queries started at different times get the same results. By looking at entire queries across countries (i.e. $query_i^{USA}$ and $query_i^{AT}$), we can understand whether identical queries started at the the same time from different locations get the same results.

The data we collected spans from 2013-10-20 06:20 UTC - 2013-10-22 22:20 UTC. Each query starts exactly on the 20-minute interval and lasts for exactly 30 minutes. In this way, we collect 194 datasets in total from each country. In the between-country case, we compare the entire dataset as both queries are running in both locations. In the between-time case, we only compare the 10-minute overlaps between $query_i^{C}$ and $query_{i+1}^C$.

\begin{table}
  \caption{Number of Comparisons, Median, Average, and Standard Deviation of Twitter ID Jaccard Scores across all comparisons. Because the temporal comparisons are between query, we have one less than in the geographic comparison.}
  \begin{tabular}{| l | c | c | c | c |}
  \hline
  Comparison & N & Median & Mean & STD \\
  \hline
  Geographic Comparison & 194 & 0.976 & 0.941 & 0.092 \\
  \hline
  USA Time Comparison & 193 & 0.996 & 0.995 & 0.003 \\
  \hline
  Austria Time Comparison & 193 & 0.996 & 0.942 & 0.186 \\
  \hline
  \hline  
  \end{tabular}
  \label{tab:queryresults}
\end{table}

\subsection{Between-Country Results}
To compare the datasets, we calculate the Jaccard score of the Tweet IDs from $query_i^{USA}$ and $query_i^{AT}$. We then take the median, average, and standard deviation of these Jaccard scores. These results are reported in the first row of Table~\ref{tab:queryresults}. Here, we see a very high average and a very low standard deviation between the Jaccard scores, indicating that the results from these two queries are nearly identical. These results indicate that no preference was given to the queries originating in the United States. We are hopeful that researchers outside of the US will obtain similar results.

\subsection{Between-Time Results}
To compare the datasets, we fix the country $C$ and calculate the Jaccard score of the Tweet IDs from $query_i^C$ and $query_{i+1}^C$. The results for the USA and Austria queries are shown in rows 2 and 3 of Table~\ref{tab:queryresults}, respectively. In the case of the USA, we see even stronger results, with an extremely high average and an extremely low standard deviation. Here, we can see that the overlapping times receive practically the same dataset in all cases. In the case of the Austrian datasets, we see that there is a wider distribution of Jaccard scores between query windows, however we continue to see an extremely high mean, which gives us confidence in the coverage in these results.

\section{Discussion and Future Work}
In this work we ask how to find bias in the Streaming API without the need for costly Firehose data. We test the representativity of another freely-available Twitter data source, the Sample API. We find that overall the tweets that come through the Sample API are a representative sample of the true activity on Twitter. We propose a solution to harness this data source to find time periods where the Streaming API is likely biased. Finally, we show that results obtained through the Streaming API for a given time period see a significant amount of overlap between queries generated from both the United States and from Austria, and between queries started at different points in time.

One question that arises is how to integrate the results of our framework into an individual's research. One solution is to focus research efforts on time periods where no (or little) bias is found in the dataset. Another potential solution is to purchase Twitter data from a reseller such as Topsy\footnote{\url{http://www.topsy.com}} or Gnip, but only paying for the biased time periods. A third solution is to incorporate other forms of social media such as blogs. This allows for a multifaceted view of the data, and can give the researcher more depth in times where their Twitter data may be of question. One way to incorporate these other views is to cross-reference users from Twitter with other social media outlets, one such solution has been proposed in~\cite{ZafaraniL13}.

We have studied the feasibility of our method with the sparse signal that we get from the Sample API. Overall, we find that this method can be used when the query issued by the researcher receives a lot of attention from Twitter users. We also find that this method is less useful when the query receives less data. One potential way to alleviate this problem is to use other bootstrapping methods such that proposed in~\cite{kirk2009gaussian}, which takes into account neighboring values to compute the values. Future work will attempt to find bias in sparse data scenarios, and adapting these methods to the speed and ephemerality of Twitter data.


\bibliographystyle{abbrv}
\bibliography{references}

\end{document}